# A Single Detect Focused YOLO Framework for Robust Mitotic Figure Detection

Yasemin Topuz[1], M. Taha Gökcan[1], Serdar Yıldız[1], and Songül Varlı[1]

[1]Department of Computer Engineering, Faculty of Electrical and Electronics Engineering, Yildiz Technical University, Davutpasa, 34220 Esenler, Istanbul, Türkiye

**Mitotic figure detection is a crucial task in computational pathology, as mitotic activity serves as a strong prognostic marker for tumor aggressiveness. However, domain variability that arises from differences in scanners, tissue types, and staining protocols poses a major challenge to the robustness of automated detection methods. In this study, we introduce SDF-YOLO (Single Detect Focused YOLO), a lightweight yet domain-robust detection framework designed specifically for small, rare targets such as mitotic figures. The model builds on YOLOv11 with task-specific modifications, including a single detection head aligned with mitotic figure scale, coordinate attention to enhance positional sensitivity, and improved cross-channel feature mixing. Experiments were conducted on three datasets that span human and canine tumors: MIDOG ++, canine cutaneous mast cell tumor (CCMCT), and canine mammary carcinoma (CMC). When submitted to the preliminary test set for the MIDOG2025 challenge, SDF-YOLO achieved an average precision (AP) of 0.799, with a precision of 0.758, a recall of 0.775, an F1 score of 0.766, and an FROC-AUC of 5.793, demonstrating both competitive accuracy and computational efficiency. These results indicate that SDF-YOLO provides a reliable and efficient framework for robust mitotic figure detection across diverse domains.**

MIDOG | Domain shift | Small object detection
Correspondence: *ytopuz@yildiz.edu.tr*

## Introduction

Mitosis is the phase in which a cell divides into two daughters while keeping the chromosome number, and loss of this control is a sign of malignant transformation and unchecked proliferation (1–4). In routine tumor pathology, the density of mitotic figures is a reliable indicator of tumor aggressiveness and is included in many grading systems. Major guidelines recommend reporting this count in pathology reports (5, 6), and higher values are consistently associated with poorer survival across disease stages (6–8). Automating key steps in the pathology workflow shortens the diagnostic process and supports consistent decisions, and early and accurate cancer diagnosis saves lives.

Counting mitotic cells is one of the most difficult and time-consuming tasks for pathologists in all types of cancer (9, 10). The challenge stems from the wide morphological variation of mitotic figures and their close resemblance to non-mitotic nuclei. This resemblance drives interobserver differences, and reports on the same case can vary with the pathologist's level of expertise (11–14). A further obstacle is the heterogeneity introduced by slide preparation across laboratories. Differences in stains, reagents, manufacturers, protocols and room conditions shift the appearance of tissue and create the well-known domain shift problem (15, 16).

Digital pathology provides the technical basis for automation. Conventional glass slides are scanned with dedicated microscopes and high quality sensors to produce whole slide images (WSIs) that capture tissue at high resolution. These capabilities enable computer aided diagnosis (CAD) systems that help pathologists work faster, and improve reproducibility (17, 18).

Domain shift remains a major obstacle. Differences in scanner models and acquisition settings such as compression, color rendering, brightness, and contrast change how mitotic figures appear. As a result, models that perform well on internal data often degrade on external test sets (19). In some studies the scanner effect is stronger than variability from slide preparation (17, 18, 20). Tissue type adds further heterogeneity and alters the visual pattern of mitosis (21). To judge generalizability, algorithms should be validated on external data and analyzed for statistical robustness. CAD systems intended for real world use need to remain sensitive to differences in scanner, tissue, staining protocol, and case mix and they should be robust on real world data.

We propose **SDF-YOLO**, a version of YOLO adapted to mitosis detection with a focus on small and rare targets and robustness under domain shift.

## Material and Methods

**A. Dataset.** To ensure broad domain coverage, our model was trained and evaluated on three mitotic figure detection datasets encompassing diverse tissue types. Among these, the MIDOG++ dataset (22) was generated by digitizing samples from four laboratories, two species, and seven tumor tissue sections using four different scanners. This dataset consists of 503 Hematoxylin and Eosin (H&E)-stained High Power Field (HPF) images representing seven tumor categories, within which expert pathologists annotated 11,937 mitotic figures. Owing to its multi-laboratory and multi-scanner origin, MIDOG++ introduces substantial domain variability. To further extend this variability, two canine tumor datasets were incorporated: the Canine Cutaneous Mast Cell Tumor (CCMCT) dataset (23), comprising 32 whole-slide images of canine mast cell tumors with exhaustive mitotic annotations totaling 44,880 mitotic figures, and the Canine Mammary Carcinoma (CMC) dataset (24), which includes 21 whole-slide images of canine breast carcinoma annotated with 13,907 mitotic figures. Together, these datasets intro-duce species- and tissue-level domain shifts, thereby complementing the human data and providing a comprehensive basis for model training and evaluation.

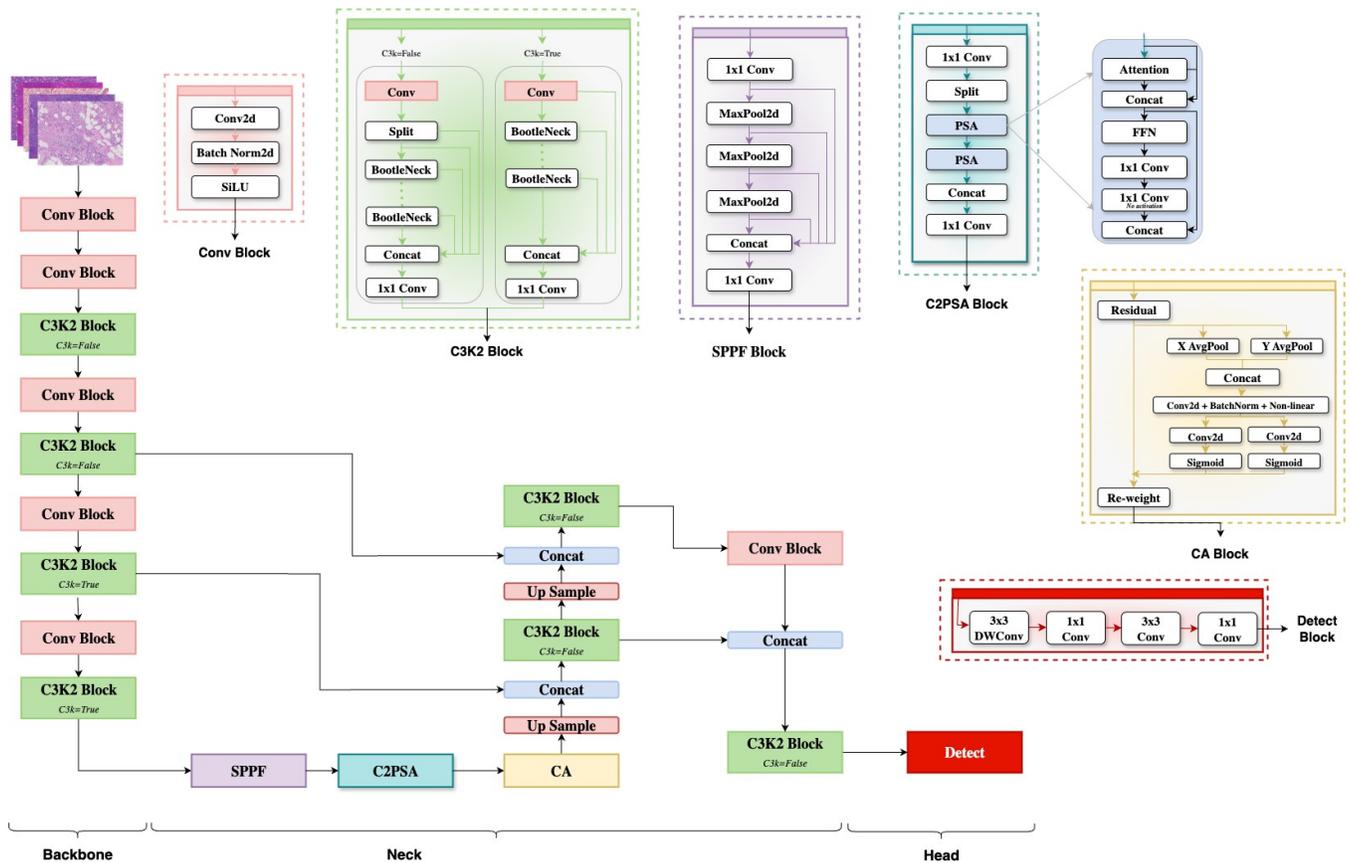

**Fig. 1.** SDF-YOLO Architecture

For mitotic cell detection, we adopted a simple and reproducible training protocol. From the MIDOG++ dataset, 503 ROI images representing seven tumor types were randomly sampled on a per-section basis to ensure balanced representation across categories and then split into 352/50/101 images for training, validation and testing, respectively. The CCMCT dataset was divided into 20/9/3 images, and the CMC dataset into 15/4/2 images, following the same train/validation/test order.

To address class imbalance, we used a custom patch sampler that enforced a 1:1 ratio between positive patches (containing at least one annotated mitotic figure) and empty patches.

**B. SDF-YOLO: proposed methodology.** We introduce Single Detect Focused YOLO (SDF-YOLO), a tailored architecture derived from YOLOv11 through a series of targeted modifications. Designed to capture the intrinsic characteristics of the mitosis detection task, SDF-YOLO achieves a substantial reduction in parameter count while maintaining robust detection capability. This compact design translates into faster inference, making the model particularly well-suited for large-scale histopathology applications. The architecture of SDF-YOLO is depicted in Figure 1.

YOLOv11 introduces efficiency-oriented blocks such as C3K2, SPPF, and C2PSA, and is reported by its authors to improve small-object accuracy while preserving real-time inference. Since mitotic figures occupy small regions, a YOLOv11-based solution is a natural fit; however, the task targets a single class and almost the same size at all times, so we simplify the design accordingly.

SDF-YOLO follows the standard backbone–neck–head layout, but with task-specific simplifications:

**Coordinate Attention (CA)**: After C2PSA, we insert Coordinate Attention (CA) to preserve positional cues along x and y and to re-weight channels using this spatial context. This helps localize small, position-sensitive targets like mitoses.

**Single Head**: Instead of the usual multi-scale (P3/P4/P5), we predict only from P4. For a 640×640 input, P4 has a stride of 16, resulting in a 40×40 grid. Since mitoses are labeled with 50×50 boxes, each object covers roughly a 3×3 area on P4—sufficient for localization without the overhead of additional scales. This reduces computation and latency while maintaining sensitivity to small objects.

**Channel mixing tweak**: YOLOv11's head combines depthwise and standard convolutions. We replace the second depthwise layer with a standard 2D convolution to improve cross-channel mixing, which stabilizes box and score regression in the single-scale setting.

By removing P3/P5 and refining the head, SDF-YOLO reduces parameters and compute, shortens training and inference, and focuses capacity where it matters most for a single, size-consistent class. CA strengthens localization of small targets, and the simplified head improves feature fusion with-out a heavy cost.

**C. Network Training.** Model training was performed over 100 epochs with a batch size of 16 and a fixed input resolution of 640×640. Stochastic gradient descent (SGD) was employed as the optimizer, configured with a learning rate of 0.01 and a weight decay of 0.0005, while all other parameters were kept at their default values.

The augmentation strategies implemented are categorized into geometric transformations and color augmentations. We focused on manipulating cell morphology through geometric transformations, employing techniques such as translation, rotation, and flipping, which do not alter cell size. To significantly enhance model robustness against variations in scanner types and stain characteristics, we applied color augmentations, including adjustments to brightness and contrast. The checkpoint used for reporting was selected by mitosis mAP@0.5 on the validation split.

All experiments were conducted using the Ubuntu 22.04 operating system. To run the experiments, 2x NVIDIA RTX 4090 GPUs and an Intel i9-13900K CPU were used as resources for computation.

## Results

Using a fixed protocol with an IoU threshold of 0.4 and a confidence threshold of 0.45, we discarded predictions smaller than 35 px and conducted inference at 640×640 with flip-only test-time augmentation. When the selected model was submitted to the Mitosis Domain Generalization Challenge 2025 (MIDOG) (25) preliminary test set and evaluated accordingly, it achieved an average precision of 0.799, with a precision of 0.758, a recall of 0.775, and an F1 score of 0.766 on the validation set. Detection and localization performance was further evidenced by a FROC-AUC of 5.793.

In addition to the overall evaluation, the performance was further analyzed across tumor domains. As illustrated in Figure 2, domain-specific results show variations in detection capability: Tumor1 achieved a precision of 0.673 and recall of 0.972, Tumor2 obtained 0.799 and 0.714, Tumor3 reached 0.719 and 0.776, while Tumor4 yielded 0.865 and 0.842, respectively. These results demonstrate that although precision remained consistently high across domains, recall varied more prominently, particularly for Tumor2.

## Discussion

This work presented SDF-YOLO, a simplified YOLO-based framework designed to address the challenges of mitotic figure detection under domain variability. Unlike conventional multi-scale detectors, SDF-YOLO adopts a single detection head aligned with the size of mitotic figures, thereby reducing computational cost while maintaining sensitivity to small and rare objects. The addition of coordinate attention further enhanced robustness to scanner- and stain-related domain shifts.

The performance evaluation on the MIDOG2025 preliminary test set confirms the effectiveness of this design. Achieving an AP of 0.799, F1-score of 0.766, and a FROC-AUC of 5.793, SDF-YOLO demonstrated competitive detection and localization performance in a rigorous benchmark setting. These results are particularly encouraging given the diversity of the training data, which included both human and canine tumor domains.

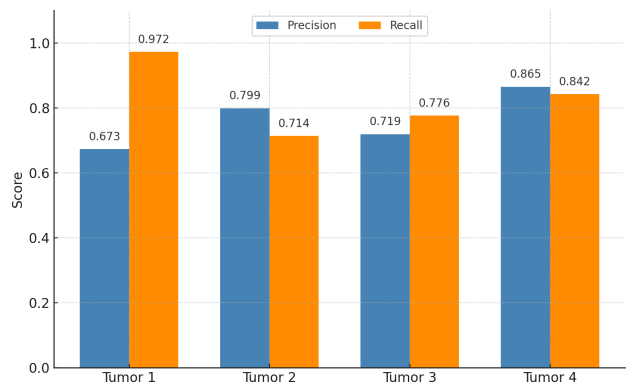

**Fig. 2.** Precision and Recall by Tumor Domain.

Despite these strengths, limitations remain. The assessment was restricted to the preliminary phase of the MIDOG2025 challenge, and further validation on the final test set is necessary to confirm generalizability. Moreover, while the single-detection strategy proved highly effective for mitotic figure detection, broader applications involving multi-class or multi-scale detection may require adapting the architecture. Overall, SDF-YOLO balances efficiency, accuracy, and robustness, establishing a strong baseline for domain-generalizable mitotic figure detection and demonstrating its practical potential in the context of international benchmarks such as MIDOG2025.